\documentclass[twocolumn,showpacs,a4,prb]{revtex4}
\usepackage{graphicx}
\usepackage{color}
\usepackage{dcolumn}
\usepackage{bm}
\begin{document}
\preprint{ODCMR.TEX}
\title{Spin order and lattice frustration in optimally doped manganites. A high temperature NMR study.}
\author{N. Panopoulos, D. Koumoulis, G. Diamantopoulos, M. Belesi, M. Fardis, M. Pissas, and G. Papavassiliou}
\affiliation{Institute of Materials Science, NCSR,  Demokritos, 153 10 Aghia Paraskevi, Athens, Greece}
\date{\today }
\begin{abstract} The physics underlying the complex glassy phenomena, which accompany the formation of polarons in optimally doped manganites (ODM) is a cumbersome issue with many unexplained aspects. In this article we present $^{139}$La and $^{55}$Mn NMR in the temperature range $80$K - $900$K of ODM La$_{0.67}$Ca$_{0.33}$MnO$_3$. We show that local lattice distortions, established in the Paramagnetic (PM) phase for $T<700$K, induce a genuine spin-glass state, which for $T<T_c$  consolidates with the Ferromagnetic (FM) state into a single thermodynamic phase. Comparative NMR experiments on La$_{0.77}$Ca$_{0.23}$MnO$_3$, La$_{0.59}$Ca$_{0.41}$MnO$_3$, and La$_{0.70}$Sr$_{0.30}$MnO$_3$ demonstrate the dominant role of lattice distortions, which appear to control (i) the stability of the spin glass phase component and (ii) the kind (1st or 2nd order) of the PM-FM phase transition. The experimental results are in agreement with the predictions of the compressible random bond - random field Ising model, where consideration of a strain field induced by lattice distortions, is shown to invoke at $T_c$ a discontinuous (1st order like) change of both the FM and the "glassy" Edwards-Anderson (EA) order parameters.
\end{abstract}
\maketitle

\section{Introduction}

The study of strong electron correlations in transition metal oxides unveiled a  complex  world of interweaving properties, concerning their spin, charge, and crystal structure. Predominant examples are high temperature superconducting cuprates and hole doped manganites. Competition among different interactions in these systems generates  spectacular phenomena, such as the formation of charge and spin stripes \cite{Mook99,Teitel'baum00,Tranquada94,Tranquada95}, mesoscopic phase separation \cite{Dagotto01,Faeth99,Mathur03}, and the colossal magnetoresistance (CMR) effect \cite{Murakami03,Mathur97}. At the same time frustration of interactions gives rise to the appearance of freezing and glassiness \cite{Schmalian00,Panagopoulos05}, expressed with slow relaxation, aging, and other signatures of glassy systems \cite{Julien01,Chatterjee02,Papavassiliou01}. However, it is not yet clear, whether this kind of glassiness is a fundamental property of strongly correlated electron systems, or the consequence  of quenched disorder, which produces uncorrelated charge and spin fluctuations.

\begin{figure}[tbp]
\centering
\includegraphics[angle=0,width=8cm]{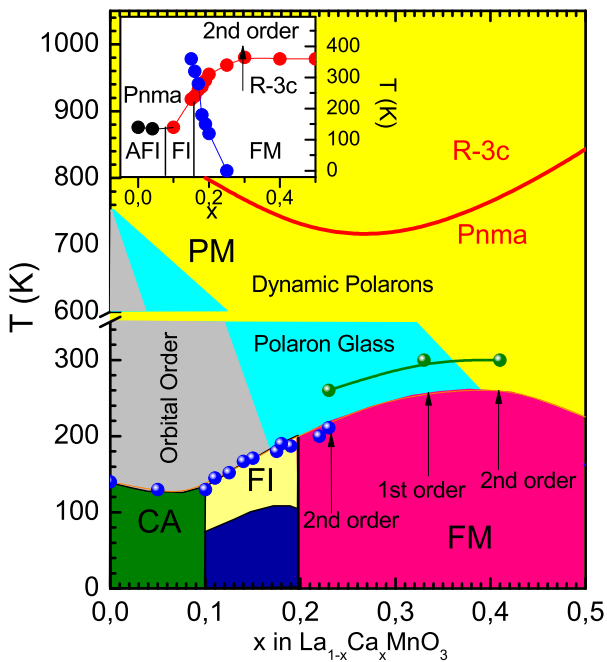}
\caption{The phase diagram of La$_{1-x}$Ca$_{x}$MnO$_3$ for $0.0\leq x\leq 0.5$. The polaron glass and dynamic polaron regimes are defined according to ref. \cite{Lynn07}. The green points (line) define the PM to FM transition line in magnetic field 9.4 Tesla. The inset shows the corresponding phase diagram for La$_{1-x}$Sr$_{x}$MnO$_3$ from ref. \cite{Asamitsu96}. The blue line is the transition line between the R-3c and Pnma crystal structures.}
\label{fig1}
\end{figure}

In the case of hole doped manganites, exemplified by the prototype La$_{1-x}$Ca$_{x}$MnO$_3$ (LCMO) family, substitution of La$^{3+}$ ions with the divalent alkaline-earth metal ion Ca$^{2+}$, invokes the replacement of Jahn-Teller (JT) active Mn$^{3+}$ ions with JT inactive Mn$^{4+}$ ions. Here, frustration is generated by competition between (i) coherent Jahn-Teller (JT) lattice  distortions, which favour charge localization, and (ii) the Double Exchange mechanism, which favours motion of $e_g$ electrons between adjacent FM ordered Mn$^{3+}$ and Mn$^{4+}$ ions, and tends to smooth out lattice distortions. In the doping range ($0.2<x<0.5$) the FM metallic phase is imposed, and the ground state is that of an isotropic FM metal (Figure \ref{fig1}). However, a number of experiments \cite{Adams00,Dai00,Zhang01} have shown that at optimally doping, $x\approx 0.33$, in the region of $T_c$, nanoscale lattice polarons (local lattice and spin textures) are formed, which drive the PM to FM phase transition to 1st order \cite{Mira99,Rivadulla04,Assaridis07}. At the same time, neutron scattering experiments, have shown that polarons might form a glassy phase for  $T>T_c$, which subsequently melts to a polaron fluid by further increasing temperature \cite{Argyriou02,Lynn07} (Figure \ref{fig1}). Other experiments revealed that the anomalous 1st order "like" phase transition is accompanied by frequency and temperature dependence \cite{Cordero01}, and strong relaxation effects \cite{Heffner00,Merithew00}, which are reminisnece of relaxor ferroelectrics and spin glasses. Similar effects have been observed in Nd$_{0.70}$Sr$_{0.30}$MnO$_3$ \cite{FernandezBaca98}. By contrast, no such effect have been observed in other ODM manganites, such as La$_{0.70}$Sr$_{0.30}$MnO$_3$ (LSMO$(0.30)$), which exhibit a conventional 2nd order phase transition \cite{Mira99,Assaridis07,Ghosh98}.

It is evident by now that the remarkable differences between the Ca-doped and Sr-doped systems  are imposed by their different high temperature crystal symmetry: In LCMO$(0.33)$, a Rhombohedral to
Orthorombic (R-3c to Pnma) phase transition occurs by cooling, at $\approx 700$K \cite{Souza08}, accompanied by the onset of strong collective
JT distortions, while LSMO$(0.33)$ remains in the Rhombohedral R-3c phase at all temperatures \cite{Asamitsu96} (inset of Figure \ref{fig1}), where static JT displacements are forbidden by the crystal symmetry.
However, it is unclear whether the observed "glassiness" in LCMO$(0.33)$ is simply due to random freezing of polaronic distortions, or a collective spin-glass transition of the Ising type, driven by frustrated interactions and internal stresses.

\begin{figure}[tbp]
\centering
\includegraphics[angle=0,width=9cm]{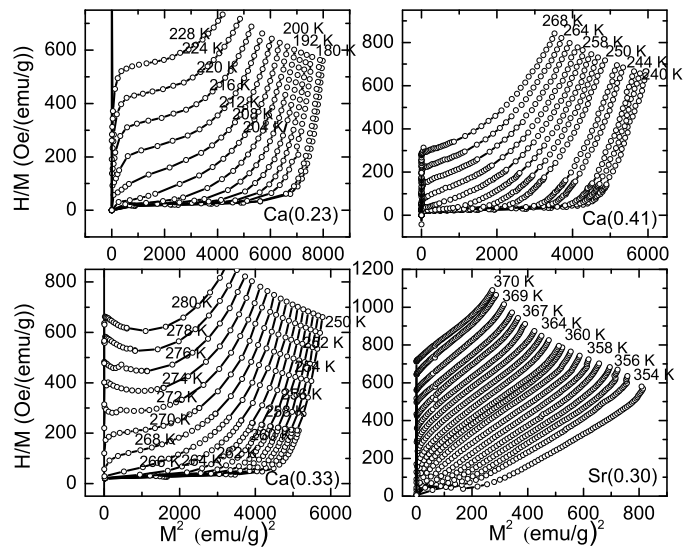}
\caption{H/M vs M$^2$ plots of isotherms in the vicinity of the Curie Temperature T$_c$ for LCMO$(0.23,0.33,0.41)$, and LSMO$(0.30)$ systems.}
\label{fig2}
\end{figure}

Another puzzling issue is that by shifting away from $x=0.33$, where the stronger glassy phenomena are observed, the 1st order PM-FM phase transition turns to 2nd order (Figure \ref{fig1}). Figure \ref{fig2} demonstrates $H/M$ vs $M^2$
isotherms of LCMO($0.23, 0.33, 0.41$) and LSMO($0.30$) in the vicinity of $T_c$. It is clearly seen that the panel for $x=0.33$ shows negative slope in the lower $M^2$ region, which according to the Banerjee criterion \cite{Mira99,Banerjee64} is a clear sign that LCMO($0.33$) belongs to the 1st order transition. By moving away from optimal doping, in both directions, the slope becomes positive signifying that the transition becomes of the 2nd order. However, in the lower doping regime, scaling arguments, which characterize 2nd order phase transitions are not directly applicable, and there are questions about the exact nature of the phase transition \cite{Rivadulla04}.

In order to shed light to this difficult issue, we have performed $^{139}$La and $^{55}$Mn NMR in the temperature range $80$ K to $900$ K, on optimally doped LCMO$(0.33)$, as well as LCMO($0.23$), LCMO($0.41$), and LSMO($0.30$) powder samples. In contrast to diffraction techniques, where long range order with a coherence length of at least $100$nm is required, NMR is a local probe and therefore ideal for the characterization of short range ordered systems. As far as we know, this is the first time that NMR experiments on manganites were performed at temperatures as high as $900$K. Our experiments suggest that in case of LCMO($0.33$) strong local lattice distortions, apparently of polaronic origin \cite{Lynn07}, are formed below the Rhombohedral to Orthorombic transition temperature at $\approx 700$K. These distortions are responsible for the appearance of a spin-glass state in the PM phase, while for $T<T_c$ the spin glass state merges with the FM phase, comprising a new thermodynamic phase (a kind of collective spin-glass). The order in this novel phase resembles the order of spin-glasses in a magnetic field, i.e. the long range FM order and the short range glassy order coexist. Contrary to reentrant spin-glass phases, by lowering temperature the growing FM order is shown to truncate the glassy phase component.

This intriguing phase transition is excellently described by the compressible random interaction-random field Ising model \cite{Papantopoulos94}, where strains induced by polaronic distortions are taken into consideration. In case of LCMO$(0.33)$ a discontinuous change of the spin and EA order parameters at $T_c$ are foreseen by the model, while for $T<T_c$ the EA order parameter $q_{EA}$ decays by decreasing temperature, in agreement with the experimental results. We notice that the EA order parameter $q_{EA}$ in the FM state is defined by $\bar q_{EA}=q_{EA}-M^2$, where $q_{EA}=\frac{1}{N}\sum_i\left<S_i\right>^2=[\left<S_i\right>^2]_{av}$. The spin operator $S_i$ takes the values $\pm 1$, $\left<...\right>$ represents the time average, and $[....]_{av}$ denotes the disorder average. Experiments show that by moving away from optimal doping, lattice distortions faint out and the glassy component weakens rapidly, while the absence of an internal strain field is responsible for turning the phase transition from 1st order to 2nd order, as predicted by the model.

\section{Materials and Methods}

La$_{1-x}$Di$_x$MnO$_3$ (Di=Ca, Sr) samples were prepared by thoroughly mixing high purity stoichiometric amounts of CaCO$_3$ (SrCO$_3$), La$_2$O$_3$, and MnO$_2$. The mixed powders formed in pastile form, reacted in air at $1400$ $^o$C for several days with intermediate grinding and then slowly cooled down to room temperature. X-ray diffraction measurements were performed on a D500 SIEMENS diffractometer, showing that all samples are single phase materials with very good crystallinity. Magnetization measurements were performed on a Quantum Design MPMSR2 superconducting quantum interference device magnetometer. At temperatures higher than $350$K, magnetization measurements were performed on a Lakeshore 3700 vibrating-sample magnetometer (VSM).
The NMR experiments were performed on a home built broadband spectrometer operating in the frequency range $5-800$ MHz, in $9.4$ Tesla and zero external magnetic fields. Spectra were acquired by the spin-echo point by point method while varying the frequency, because of the large spectral width of the resonance lines. An Oxford 1200CF continuous flow cryostat was employed for measurements in the temperature range $80-350$ K and an Oxford HT1000V furnace for measurements in the range $300-900$ K.

\section{High Temperature $^{139}La$ NMR}

The local magnetic and structural properties of the PM phase for all samples have been investigated  by applying $^{139}$La NMR in magnetic field $9.4$ Tesla, at temperatures as high as $900$K.
In the presence of an external magnetic field $B$, $^{139}$La($I=7/2$) nuclei experience the Zeeman interaction, which splits the degenerate nuclear energy levels into $2I+1$ equidistant energy levels, with energies $E_m=m\gamma \hbar B$. In addition, the $^{139}$La nucleus is coupled to the local electric field gradient (EFG) tensor, through its electric quadrupole moment $Q$. We notice that a site with cubic symmetry has an EFG equal to zero; so the size of the effect of the electric quadrupole interaction on the NMR spectrum measures the degree of deviation from cubic symmetry of the surroundings of the nuclear site.
In the presence of the Zeeman and Quadrupolar interactions the frequency of the NMR spectrum corresponding to transitions between levels $m$ and $m-1$ is given by,
$\omega_m=\gamma B + \frac{3}{2}\frac{e^2qQ}{4\hbar I\left(2I-1\right)}\left(3cos^2\theta -1-\eta sin^2\theta cos^2\phi \right)\left(2m-1\right)$.
Here, $e$ is the charge of the electron, $eq$ is equal to $V_{zz}$, $\eta$ is the asymmetry parameter of the EFG tensor $\eta=\frac{\mid  V_{yy}-V_{xx}\mid }{V_{zz}}$, and $\theta$ , $\phi $ are the angles between the principal axis of the EFG tensor and the magnetic field.
In case of LCMO manganites, the magnetic field $B$ at the site of the La nuclei is equal to the sum of the external field $B$ and the transferred hyperfine field ${B_{hf}=(1/\gamma \hbar )A<S>}$, where ${A}$ is the hyperfine coupling constant and ${<S>}$ the average electronic spin of the eight nearest Mn neighbors \cite{Papavassiliou97, Papavassiliou99}. In the PM phase, large frequency shifts proportional to the magnetic susceptibility are produced by the hyperfine coupling in the presence of an external magnetic field.
According to the above equation, the transition between the $m=+1/2$ and $m=-1/2$ levels (the so called central transition) is unaffected by the electric quadrupole interaction to 1st order, while the distance in frequency of all other transitions (sattelites) from the central transition, depends solely on the quadrupolar interaction. In a powder sample, the crystal axes, and hence the EFG, are distributed at random angles with respect to the applied magnetic field. The angular average of the satellite patterns gives rise to a characteristic frequency distribution. The size of this distribution is proportional to the EFG (and consequently to the lattice distortions) and independent of the magnetic field.

\begin{figure}[tbp]
\centering
\includegraphics[angle=0,width=8cm]{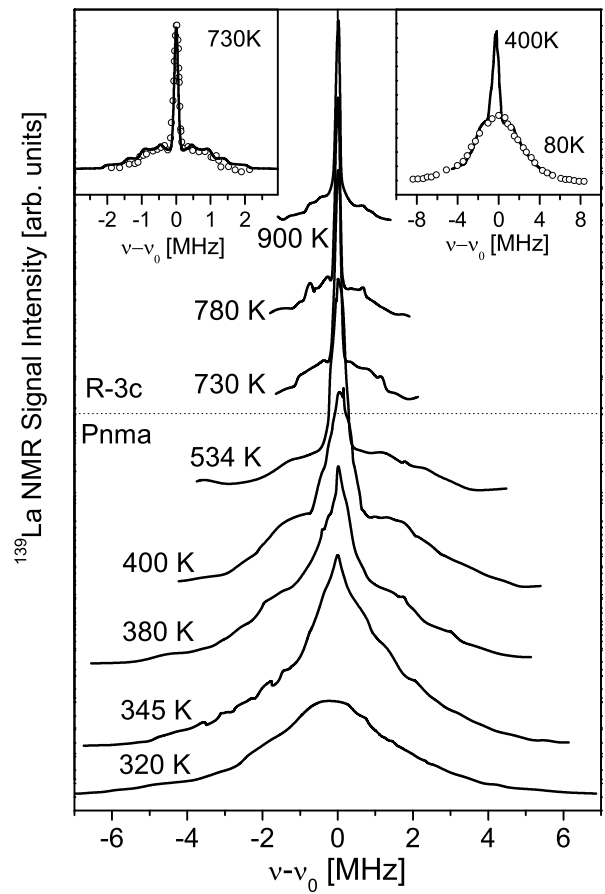}
\caption{$^{139}$La NMR spectra for optimally doped LCMO($0.33$), in $9.4$ Tesla external magnetic field, as a function of temperature. Spectra are presented relatively to the central line frequency $\nu _0$. The dotted line shows schematically the border line between the R-3c and Pnma phase regimes. The left inset depicts the calculated powder pattern spectrum in comparison to the experimental spectrum at $T=730$K. The right inset shows a comparison of spectra at $80$K (open circles) and $400$K (solid line).}
\label{fig3}
\end{figure}

Figure \ref{fig3} demonstrates $^{139}$La NMR spectra of the LCMO($0.33$) system in the temperature range $320-900$K. At temperatures higher than $700$K spectra exhibit typical powder patterns, consisting of the narrow central line ($\approx 150$kHz broad), and the broad satellite frequency distribution (SFD) ($\approx 1.6$MHz broad); the latter is the fingerprint of the EFG in the $R-3c$ Rhombohedral phase. The inset at the top left side of Figure  \ref{fig3} depicts the calculated powder pattern spectrum at $730$K, which matches excellently with the experimental spectrum. For the simulation an electric quadrupolar coupling $\nu _{Q}=\frac{e^2qQ}{h}=21$MHz, asymmetry parameter $n=0.5$ and a dipole-dipole interaction of the La nuclear sites with the Mn ions of $\approx 20$kHz were considered. By decreasing temperature, at $T\approx 700$K, a significant increase of the SFD width is observed, which marks the transition from the $R-3c$ to the $Pnma$ crystal structure \cite{Souza08}, and the appearence of strong incoherent JT displacements. This structural phase transformation is shown by synchrotron radiation measurements to be independent of the applied magnetic field \cite{Souza08}. Attempts to simulate the powder pattern of the sattelites for $T<700$K were only possible by assuming a distribution of EFG tensor components, i.e. a distribution of lattice distortions. Most remarkably, by further decreasing temperature, the average width of the SFD remains invariant down to the lowest measured temperature, as clearly seen in the inset at the top right side of Figure  \ref{fig3}. Hence, from the viewpoint of the La site, the size and distribution of lattice distortions in the Pnma phase does not change by decreasing temperature. At the same time, the narrow central future starts to broaden by cooling and disappears at $\approx 320$K, in agreement with previous measurements \cite{Sakaie99}. The extreme broadening of the central line, which up to 1st order depends solely to magnetic interactions, is a clear sign about the onset of a broad distribution of magnetic susceptibilities in the $Pnma$ phase (the NMR signal broadening is proportional to the product of the magnetization and the susceptibility distribution width). Evidently, the appearance of short range correlated polarons \cite{Lynn07}, produces a distribution of exchange couplings, which is directly reflected on the spin order of the system.

\begin{figure}[tbp]
\centering
\includegraphics[angle=0,width=8cm]{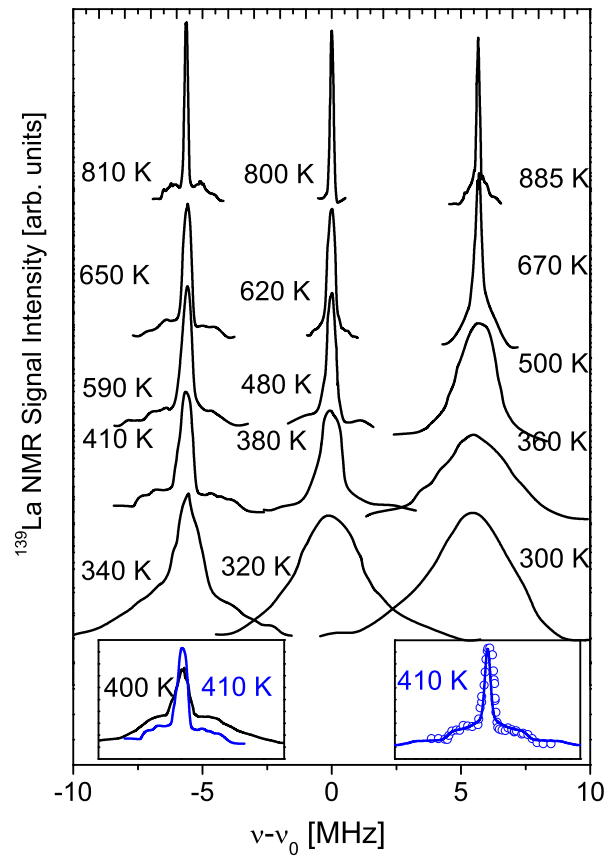}
\caption{$^{139}$La NMR spectra for LCMO, $x=0.23$(left), $0.41$ (middle), and LSMO $x=0.3$ (right) at various temperatures. Spectra are presented relatively to the central line frequency $\nu _0$. The left inset shows a comparison of LCMO $x=0.23$ (blue), and $0.33$ (black) experimental spectra at $T=410$K, and $400$K respectively. The right inset depicts the calculated powder pattern spectrum in comparison to the experimental spectrum for $x=0.23$ at $T=410$K.}
\label{fig4}
\end{figure}

\begin{figure}[tbp]
\centering
\includegraphics[angle=0,width=8cm]{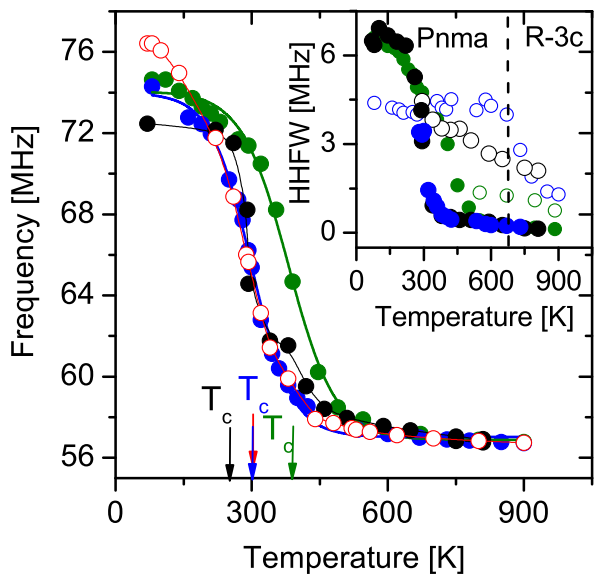}
\caption{The $^{139}$La NMR frequency as a function of temperature for LCMO$(0.23)$ ($\bullet $), LCMO($0.33$) ($\textcolor{blue}{\bullet }$), LCMO($0.41$) ($\textcolor{red}{\circ }$), and LSMO$(0.30)$ ($\textcolor{green}{\bullet }$). In the inset the width of the NMR frequency distribution is shown for (i) LCMO($0.23$) central line ($\bullet $), and satellite powder pattern ($\circ $), (ii) LCMO($0.33$) central line ($\textcolor{blue}{\bullet }$), and satellite powder pattern ($\textcolor{blue}{\circ }$), and (iii) for LSMO($0.30$) central line ($\textcolor{green}{\bullet }$), and satellite powder pattern ($\textcolor{green}{\circ }$).}
\label{fig5}
\end{figure}

Figure \ref{fig4} shows $^{139}$La NMR spectra of the LCMO($0.23$), LCMO($0.41$), and LSMO($0.30$) systems in the temperature range $300-885$K. In case of LCMO($0.23$), at high temperatures spectra depict the typical NMR powder pattern as in LCMO($0.33$). However, two basic differences between the two systems are that (i) the SFD of LCMO($0.23$) is at all measured temperatures sufficiently narrower than the SFD of LCMO($0.33$), and (ii) the SFD width of LCMO($0.23$) varies smoothly by decreasing temperature in the PM phase, while in LCMO($0.33$), after increasing abruptly at $T\approx 700$K, it remains invariant by lowering temperature (inset in Figure \ref{fig5}). At first sight this is a surprising result, because LCMO($0.23$) exhibits (i) transition from the $Pnma$ to the $R-3c$ phase at approximately the same temperature as LCMO$(0.33)$, and (ii) neutron scattering PDF measurements have shown that local JT distortions decrease by increasing Ca doping \cite{Bozin07}. A possible explanation is that close to doping $x=0.25$, small polarons might form an ordered lattice, where Mn$^{4+}$ ions have 6 neighbouring JT distorted Mn$^{+3}$ sites, fitting together in a space filling 3D network, which minimizes lattice distortions and strains \cite{Billinge00}. The picture of reduced lattice distortions in the $Pnma$ phase of LCMO($0.23$) is supported by the caculated powder pattern spectra. The right inset in Figure \ref{fig4} depicts the calculated spectrum in comparison to the experimental one at $410$K. For the fit a quadrupolar coupling constant $\nu _{Q}=28$MHz, an asymmetry parameter $\eta=0.6$ and a dipole-dipole interaction of the La nuclear sites with the Mn ions $\approx 50$kHz were considered. In addition, a large Gaussian broadening was necessary for the fit, which is indicative of the presence of substantial lattice distortions. Nevertheless, these distortions are sufficiently weaker than in LCMO($0.33$), where in the $Pnma$ phase adequate fit is only possible by considering overlapping spectra with a broad distribution of quadrupolar coupling constants.
On the other hand, in case of LCMO$(0.41)$ the SFD is almost absent and only the magnetic broadening of the central line is observed by decreasing temperature. This result is in agreement with the phase diagram in Figure \ref{fig1}, which shows no static polarons for $x=0.41$, but rather the presence of dynamic polarons \cite{Lynn07,Argyriou02} with a zero time average of local lattice distortions in the time scale of NMR experiments. Similar spectra but for different reasons are shown by LSMO$(0.30)$. This system remains at all temperatures in the $R-3c$ Rhombohedral phase (inset in Figure \ref{fig1}), where static JT distortions are by symmetry forbidden, and therefore it depicts only a very narrow SFD, while the central line broadens gradually by lowering temperature.

Figure \ref{fig5} shows the $^{139}$La NMR frequency at $9.4$ Tesla as a function of temperature, for all four measured systems. The critical temperatures $T_c$, defined as the inflection point on  the $\nu$ vs. $T$ curves, are found to be $260$K for LCMO($0.23$), $300$K for LCMO($0.33$) and LCMO($0.41$) and $400$K for LSMO($0.30$). These values (shown as green points in Figure \ref{fig1}) define the PM to FM transition line at $9.4$ Tesla in the $T-x$ phase diagram, which is parallel shifted to higher temperatures from the transition line in zero external magnetic field. We notice that although in 2nd order phase transitions critical temperature does not exist in the presence of a magnetic field, the observed temperature variation of the NMR frequency is fairly steep, which is reminicence of a true phase transition. Furthermore, it is observed that the transfered hyperfine field at saturation, expressed by the $^{139}$La NMR frequency at $80$K, depends almost linearly on the doping concentration, in agreement with previous $^{139}$La measurements in zero external magnetic field \cite{Papavassiliou99}. The inset in Figure \ref{fig5} shows the full width at half maximum (FWHM) of the central line (filled circles) and the FWHM of the SFD (open circles) for LCMO($0.23$), LCMO($0.33$), and LSMO($0.30$). As previously discussed, in case of LCMO($0.33$) the transition from the $R-3c$ to the $Pnma$ phase by cooling, is clearly seen to occur at $T\approx 700$K. Below that temperature, the abrupt increase of the SFD width declares the appearance of strong static local JT distortions, which remain invariant down to the lowest measured temperature. It is also clearly seen that the SFD width and thus the local lattice distortions, are in LCMO($0.33$) by far the largest among the four measured systems. Another important notification is that contrary to LCMO($0.33$) where the central line disappears at $T\approx 300$K, apparently due to extreme inhomogeneous broadening, in the rest three systems the central line remains visible at all temperatures, and its width varies with temperature in exactly the same way as the $^{139}$La NMR frequency. This is a clear sign that - unlike to the LCMO($0.33$) system - there is no broad distribution of magnetic susceptibilities.

\section{$^{55}Mn$ NMR measurements in zero External Magnetic Field}

The invisible with $^{139}$La NMR "magnetic" glassy phase component for LCMO($0.33$), is possible to monitor by applying $^{55}$Mn NMR in zero external magnetic field. For reasons of comparison NMR data are compared with the prototype (in respect to the $2^{nd}$ order nature of the phase transition) LSMO($0.30$) system in the temperature range $80$K up to temperatures close to $T_c$.

\begin{figure}[tbp]
\centering
\includegraphics[angle=0,width=8cm]{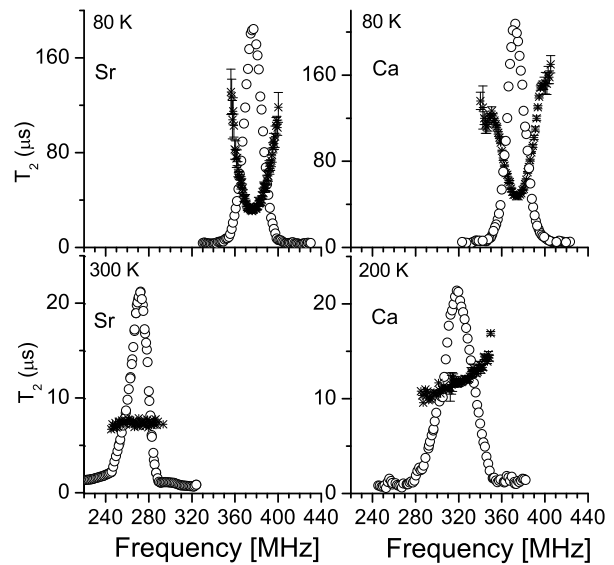}
\caption{$^{55}$Mn NMR spectra (open circles) of LCMO($0.33$) at $80$K and $200$K and LSMO($0.30$) (left figures) at $80$K and $300$K, respectively. The spin-spin $T_2$ relaxation times (crosses) as a function of frequency are also shown in the same Figures. By increasing temperature spectra shift to lower frequencies, while for LCMO($0.33$) they become asymmetric broader on approaching $T_c$ from below. The  $T_2$ minimum at the peak of the spectra at $80$K is due to the Shul-Nakamura inetractions.}
\label{fig6}
\end{figure}

$^{55}$Mn NMR in zero external magnetic field probes directly the electron spin state of single Mn ions through the hyperfine field ${B_{hf}=(1/\gamma \hbar )A<S>}$, and therefore it is possible to resolve the different Mn charge states, i.e., localized Mn$^{3+}$, Mn$^{4+}$ and the intermediate FM valence states. Figure \ref{fig6} shows $^{55}$Mn NMR spectra for both samples at $80$K and $200$K ($300$K), respectively. Spectra, as expected, were found to consist of a single line that is caused by motional narrowing, due to the fast electron hoping between the Mn$^{3+}$, Mn$^{4+}$ manganese ions with frequency much higher than the NMR frequency \cite{Papavassiliou00}. In the same figure the nuclear spin-spin relaxation time $T_2$ as  function of frequency is demonstrated. $T_2$ in manganites depends strongly on temperature \cite{Savosta99}, apparently due to fluctuations of the hyperfine field caused by hoping of the electron holes. At lower temperatures, an additional contribution to $T_2$ is observed, due to the Suhl-Nakamura relaxation mechanism \cite{Davis74}, which produces a characteristic $T_2$ minimum at the center of the spectra, in agreement with previous results \cite{Savosta01}. This frequency dependence of $T_2$ disappears by increasing temperature. In case of LSMO($0.30$) $T_2$ is shown to be frequency independent on approaching $T_c$ from below. However, in LCMO($0.33$) another frequency dependent spin-dynamics appears, which is expressed as a monotonic increase of $T_2$ by increasing frequency. This effect indicates the presence of a distribution of low frequency fluctuations close to $T_c$ in LCMO($0.33$), which is absent in LSMO($0.30$). At the same time, for $T>150$K a significant asymmetric broadening of the lineshape is observed in LCMO($0.33$), which is absent in LSMO($0.30$), in agreement with previous NMR works \cite{Papavassiliou00,Savosta01}.

\begin{figure}[tbp]
\centering
\includegraphics[angle=0,width=8cm]{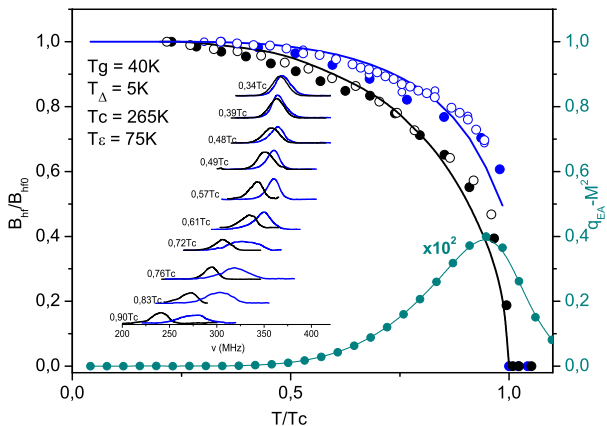}
\caption{The hyperfine filed $B_{hf}$ as a function of temperature, obtained from $^{55}$Mn NMR measurements, for LCMO$(0.33)$ ($\textcolor{blue}{\circ }$) and LSMO$(0.30)$ ($\circ$). The filled circles ($\textcolor{blue}{\bullet }, \bullet $) are the corresponding $B_{hf}$ values as obtained by Moessbauer spectroscopy on lightly $Sn$-doped samples from ref. \cite{Assaridis07}. The solid lines are theoretical fits by applying the self consistent equations (2) and (3). The inset shows the corresponding spectra for LCMO$(0.33)$ (blue lines) and LSMO$(0.30)$ (black lines).}
\label{fig7}
\end{figure}

Figure \ref{fig7} demonstrates lineshapes and the normalized hyperfine field $B_{hf}/B_{hf}(0)$ for both LCMO($0.33$) and LSMO($0.30$) as a function of temperature. For reasons of comparison the $B_{hf}/B_{hf}(0)$ values obtained with Moessbauer spectroscopy from ref. \cite{Assaridis07} are also presented in the same figure. An excellent match between the NMR and the Moessbauer hyperfine fields is observed. Another important experimental feature is that in case of LCMO($0.33$) $B_{hf}$ decreases abruptly to zero at $T_c$, a clear mark of the 1st order nature of the phase transition. On the contrary, in case of LSMO($0.30$) the hyperfine field decreases continously to zero by approaching $T_c$ from below, as expected for a 2nd order phase transition.

\section{The compressible Random-Interactions and Random Fields Ising Model}

The discontinuous  hyperfine field variation at $T_c$ and the inhomogeneous $^{55}$Mn NMR lineshape broadening for $T<T_c$ observed in LCMO$(0.33)$, are nicely explained by considering an Ising Hamiltonian, where random-interactions and random-fields, together with a strain field induced by strong local lattice distortions are taken into consideration \cite{Blinc89,Papantopoulos94}.
\begin{eqnarray}
{\cal H}=\frac{1}{2}\sum_{i,j}\left[J^0_{ij}+\sum_vJ^v_{ij}\epsilon _v\right]S_iS_j-\sum_if_iS_i\nonumber\\
+\frac{1}{2}\sum_v\tilde C_v\epsilon ^2_v.
\end{eqnarray}
The last term expresses the macroscopic elastic energy. The random interactions $J^0_{ij}$ and random fields $f_i$ are assumed to be independently distributed according to Gaussian probability densities. This is an extension of the Sherrington-Kirkpatrick model \cite{Binder86}, which in the weak-disorder limit predicts coexistence of spin-glass and FM order \cite{Binder86,Papantopoulos94,Blinc89}.

In the replica-symmetric mean field approximation \cite{Binder86}, the reduced magnetization and the EA order parameters are possible to be calculated by the self-consistent equations \cite{Papantopoulos94,Blinc89},
\begin{eqnarray}
M=\frac{1}{\sqrt {2\pi}} \int_{-\infty}^{+\infty}dz \exp\left(-z^2/2\right) \nonumber\\
\tanh\left[\left(q_{EA}\frac {T_g^2}{T^2} + \frac {T_{\Delta}^2}{T^2}\right)^{1/2}z + \frac {T_c}{T}M + \frac{T_\epsilon}{T}M^3\right]
\end{eqnarray}
\begin{eqnarray}
q_{EA}=\frac{1}{\sqrt {2\pi}} \int_{-\infty}^{+\infty}dz \exp\left(-z^2/2\right)\nonumber\\
\tanh^2\left[\left(q_{EA}\frac {T_g^2}{T^2} + \frac {T_{\Delta}^2}{T^2}\right)^{1/2}z + \frac {T_c}{T}M + \frac{T_\epsilon}{T}M^3\right]
\end{eqnarray}
where, $T_g$, $T_\Delta $ and $T_\epsilon $ are temperatures, which characterize the variance of the random interactions and random fields distributions, and the strength of the strain field, respectively \cite{Papantopoulos94}.
By solving these equations self-consistently, it is possible to fit the experimental $B_{hf}/B_{hf0}$ vs. T curves in Figure \ref{fig7}. In case of LCMO$(0.33)$ data are fitted by considering $T_c=265$K, $T_g=40$K, $T_\Delta=5$K, and $T_\epsilon=75$K. This solution has a jump at $T_c$, in agreement with the experimental data. On the contrary, LSMO$(0.30)$ is nicely fitted with a 2nd order phase transition, by ignoring both randomness and strain field. For $T>T_c$ eq. 3 implies that $q_{EA}=\frac{T_\Delta ^2}{T^2}$, i.e. the EA order parameter appears to decay by increasing temperature. This is in agreement with the $^{139}La$ NMR results presented in Figure \ref{fig3}, where the appearance and gradual narrowing of the central-transition NMR line  at temperatures higher than $320$K, indicates the supression of the spin-glass phase component at elevated temperatures. In a similar way, the effective EA order parameter $\bar q_{EA}$ is shown to decay rapidly in the FM phase by decreasing temperature (Figure \ref{fig7}).

\begin{figure}[tbp]
\centering
\includegraphics[angle=0,width=8cm]{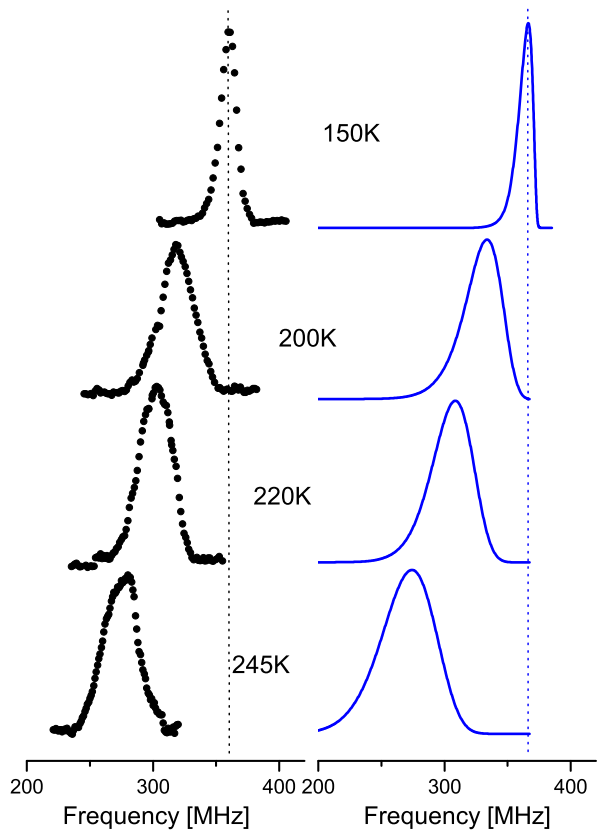}
\caption{Experimental $^{55}$Mn NMR lineshapes (left) and theoretical simulations (right) for LCMO $x=0.33$ at various temperatures, as obtained by formula (4).}
\label{fig8}
\end{figure}

On the basis of the fitting parameters, it is posible to calculate the temperature variation of the frequency distribution, $f(\nu )$, which is proportional to the probability distribution of the local magnetization $W(m)$ \cite{Blinc89},
\begin{widetext}
\begin{eqnarray}
f(\nu )\propto W(m)=\frac{1}{\left(2\pi \left[q_{EA}\frac {T_g^2}{T^2} + \frac {T_{\Delta}^2}{T^2}\right]\right)^{1/2}}\frac{1}{1-m^2} \exp\left[-\frac {\left (arctanh(m)-\frac{T_c}{T}M-\frac {T_\epsilon}{T}M^3 \right)^2}{2\left(q_{EA}\frac {T_g^2}{T^2} + \frac {T_{\Delta}^2}{T^2}\right)} \right]
\end{eqnarray}
\end{widetext}

By using eq. 4, and the fitting parameters from Figure \ref{fig7} an excellent agreement is obtained between the experimental and the simulated LCMO($0.33$) lineshapes, as shown in Figure \ref{fig8}. Evidently, the $^{55}$Mn NMR lineshape broadening observed in LCMO($0.33$) by approaching $T_c$ from below, is a sign of coexistence of long range FM order and short range spin-glass order, which accompanies the appearance of polaronic lattice distortions in this temperature region.

\section{Conclusions}

On the basis of our experimental results we anticipate that optimally doped LCMO($0.33$) shows a unique magnetic behaviour in comparison to all other measured systems, by varying the temperature across $T_c$. Specifically, by entering the Orthorombic $P_{nma}$ crystal structure at $T\approx 700$K on cooling, a broad distribution of strong lattice distortions appears, apparently of polaronic origin \cite{Lynn07}, which in turn induces an equally broad inhomogeneous distribution of Mn electron spin polarizations. Short range correlations of such polaronic distortions \cite{Argyriou02,Lynn07} appear to give rise, to a spin-glass state, which for $T<T_c$ coexists with the FM order, but it gradually disappears by further decreasing temperature. We emphasize that this genuine "collective" spin-glass state is not produced by quenched disorder, but is self generated and depends only on the competition among generic interactions such as magnetic exchange and JT interactions. Important role is played by the magnitude of lattice frustration, which creates an inhomogeneous strain field that controls the stability of the glassy state and the kind of the magnetic phase transition (1st or 2nd order). It is interesting to note that in the simplified model Ising Hamiltonian we have used, simulations show that the induced randomness on effective $J$-couplings dominates in comparison to random fields created by trapped polaronic charge carriers. Indeed, changes of the $J$ couplings (produced by local variations of the $Mn-O$ bond lengths) are expected to have a sufficiently stronger effect than the weak random dipolar fields.

\end{document}